\documentclass[conference,a4paper]{APSIPA2026}
\usepackage{amsmath}
\usepackage{graphicx}
\usepackage{multirow}
\usepackage{threeparttable}
\usepackage[backend=biber,style=ieee,]{biblatex}
\usepackage{threeparttable}

\usepackage{geometry}
\usepackage{fancyhdr}

\fancypagestyle{firststyle}{
  \fancyhf{}
  \fancyhead[C]{2026 Asia Pacific Signal and Information Processing Association Annual Summit and Conference (APSIPA ASC)}
}

\begin{document}

\title{RADAR Challenge 2026: Robust Audio Deepfake Recognition under Media Transformations}

\author{
\authorblockN{
Hieu-Thi Luong\authorrefmark{1}, Xuechen Liu\authorrefmark{2}, Ivan Kukanov\authorrefmark{3}, Zheng Xin Chai\authorrefmark{3} and
Kong Aik Lee\authorrefmark{4}
}
\authorblockA{
\authorrefmark{1} Fortemedia, Singapore \quad
\authorrefmark{2} Xi'an Jiaotong-Liverpool University, China \\
\authorrefmark{3} KLASS Engineering \& Solutions, Singapore \quad
\authorrefmark{4} The Hong Kong Polytechnic University, Hong Kong SAR, China
}

\authorblockA{
radar-challenge@hieuthi.com (Corresponding author), Xuechen.Liu@xjtlu.edu.cn, \\
\{ivan.kukanov, zhengxin.chai\}@klasses.com.sg, kong-aik.lee@polyu.edu.hk
}

}

\maketitle
\thispagestyle{firststyle}
\pagestyle{empty}
\begin{abstract}
RADAR Challenge 2026 is an APSIPA Grand Challenge on Robust Audio Deepfake Recognition under Media Transformations, designed to simulate realistic media conditions in real-world audio distribution pipelines, including compression, resampling, noise, and reverberation. It consists of two phases: an English development phase with labeled data for analysis and paper writing, and a multilingual evaluation phase containing more than 100,000 utterances in English, Singapore English, Mandarin Chinese, Taiwanese Mandarin, Japanese, and Vietnamese. Systems are evaluated using equal error rate (EER) for binary real/fake classification. This paper describes the challenge task, the construction of the data set, the evaluation protocol, and the overall results. During the challenge, 33 teams submitted to the development phase and 22 teams submitted to the final evaluation phase. The reported results highlight the remaining challenges of robust audio deepfake detection under multilingual and media-transformed conditions.
\end{abstract}

\section{Introduction}
Audio deepfake detection has become an increasingly important research topic due to rapid advances in neural speech synthesis and voice conversion systems. Modern synthetic speech can achieve highly natural quality and can be used in beneficial applications such as accessibility and conversational agents, but it also introduces risks related to impersonation, misinformation, fraud, and attacks on speaker verification systems. To address these concerns, several community benchmarks and challenges, including the ASVspoof~\cite{asvspoof2019database, asvspoof2021_summary, asvspoof5} series and recent audio deepfake detection evaluations \cite{kirill2025safe, add2022, add2023}, have driven major progress in spoofing countermeasures and evaluation methodologies. However, despite strong performance on many benchmark data sets, robustness under realistic media conditions remains a major open challenge.

In practical deployment scenarios, speech is often compressed~\cite{asvspoof5}, resampled, mixed with noise~\cite{add2023}, sound effect or background music~\cite{gao2025perturbed}, reverberated~\cite{luong2024room}, edited, or processed through proprietary and real-world media pipelines prior to analysis~\cite{muller2025deepen,kshirsagar2026investigating}. These transformations can suppress or distort spoofing artifacts while also introducing non-spoof artifacts that affect detector behavior. Existing systems may therefore achieve strong performance under clean conditions while remaining vulnerable to distribution shifts caused by media transformations, multilingual conditions \cite{muller2024mlaad,wu2025sea}, unseen spoofing sources, and limited development time constraints during challenge participation.

To study these issues, the RADAR Challenge 2026 focuses on Robust Audio Deepfake Recognition under Media Transformations\footnote{https://radar-challenge.github.io/}. Unlike benchmarks centered mainly on clean synthetic speech, we emphasize robustness under realistic media processing conditions. The challenge consists of two phases: an English development phase with labeled data for system analysis and paper writing, and a multilingual evaluation phase containing more than 100,000 utterances in English, Singapore English, Mandarin Chinese, Taiwanese Mandarin, Japanese and Vietnamese.
This paper presents the challenge task, dataset construction process, media transformation pipeline, evaluation protocol, and overall results. During the challenge, 33 teams submitted systems to the development phase and 22 teams participated in the evaluation phase. The results highlight the continuing challenges of robust audio deepfake detection under multilingual and media-transformed conditions, including transformation mismatch, spoof-source variability, cross-language generalization, and robustness under practical development constraints.

\begin{table*}[t]
\centering
\caption{Language distribution of the evaluation dataset.}
\label{tab:language_distribution}
\begin{tabular}{llrrrrr}
\hline
\textbf{Language} & \textbf{Code} & \textbf{Bonafide} & \textbf{Spoof} & \textbf{Total} & \textbf{Spoof \%} & \textbf{Dataset \%} \\
\hline
English            & en    & 15,000 & 17,923 & 32,923 & 54.44\% & 32.05\% \\
Singapore English  & en-sg &  3,500 &  6,711 & 10,211 & 65.72\% &  9.94\% \\
Mandarin           & zh    & 11,500 & 11,234 & 22,734 & 49.41\% & 22.13\% \\
Taiwanese Mandarin & zh-tw &  3,500 &  1,678 &  5,178 & 32.41\% &  5.04\% \\
Japanese           & ja    &  9,500 &  8,742 & 18,242 & 47.92\% & 17.76\% \\
Vietnamese         & vi    &  7,000 &  6,438 & 13,438 & 47.91\% & 13.08\% \\
\hline
Total              & ---   & 50,000 & 52,726 & 102,726 & 51.33\% & 100.00\% \\
\hline
\end{tabular}
\end{table*}

\section{Challenge Overview}
\textbf{Task Definition:}
The RADAR 2026 challenge focuses on binary real/fake classification for audio deepfake detection. Each participating team is provided with development and evaluation speech datasets and required to submit a score for each utterance indicating whether it is bona fide or spoofed.

\textbf{Metrics:}
System performance is evaluated using Equal Error Rate (EER), where lower values indicate better detection performance. We adopt EER as the primary metric for simplicity and consistency with prior audio deepfake detection benchmarks. Final leaderboard rankings are determined solely from the Phase 2 evaluation results.

\textbf{Challenge Motivation:}
Unlike previous audio deepfake detection challenges that primarily evaluate systems on relatively clean or lightly processed speech, the RADAR 2026 challenge emphasizes robustness under realistic media-transformed conditions. In practical deployment scenarios, speech recordings are often affected by codec compression, resampling, bandwidth limitation, additive noise, background music, reverberation, loudness normalization, dynamic range compression, trimming, and other distortions introduced during recording, transmission, editing, or social media sharing. To address this gap, the challenge evaluates both bona fide and spoofed speech under diverse and unseen media transformations, with a particular focus on robustness and generalization.

\textbf{Training Policy:}
We adopt an open-training policy, where participants are free to construct their own training pipelines using publicly accessible data, provided that data overlapping with the challenge development set are excluded. This setting better reflects practical deployment scenarios and encourages methodological diversity in multilingual coverage, augmentation strategies, pretrained models, calibration methods, score fusion, and training data selection, while maintaining fixed development and evaluation sets for fair comparison. It also allows teams to leverage their individual expertise and resources, resulting in a diverse range of approaches from different institutions and regions worldwide.

\textbf{Phase~1~(Development):}
supports system analysis, error inspection, hyperparameter tuning, and challenge paper preparation. The development set consists of English speech derived from the LlamaPartialSpoof dataset~\cite{llamapartialspof} with additional media transformations applied to both bona fide and spoofed utterances. Phase~1 was conducted over 11 days, during which each team was allowed up to two submissions per day. More than 60 teams registered for the challenge, over 30 teams submitted systems to Phase~1, resulting in 182 total submissions, and 25 teams qualified for Phase~2. Ground-truth labels were released immediately afterward to support post-hoc analysis.

\textbf{Phase~2~(Evaluation):}
assesses robustness and generalization under multilingual and media-transformed conditions. The evaluation set contains more than 100,000 utterances spanning multiple language conditions. Compared with Phase~1, the evaluation phase introduces substantially greater variability in spoofing systems and media transformation combinations. During the challenge, participants were informed only that the evaluation data included four languages, without disclosure of the regional dialect variations. Phase~2 was conducted over 8 days, during which each team was allowed one submission per day, resulting in 110 submissions. Among the 25 qualified teams, 20 submitted final evaluation scores, along with one non-ranking guest team.

\section{Dataset Construction}
\subsection{Development Set}

The development set is derived from the LlamaPartialSpoof~\cite{llamapartialspof} full-fake subset with additional media transformations applied to both bona fide and spoofed speech. In total, it contains 44,034 utterances, including 10,573 bona fide sourced from the LibriTTS~\cite{zen2019libritts} development and test partitions and 33,461 synthetic utterances generated using six speech synthesis systems: LJ JETS~\cite{lim22_lj_jets}, YourTTS~\cite{casanova22a_yourtts}, XTTS-v2~\cite{casanova24_xtts2}, GPT-SoVITS\footnote{https://github.com/RVC-Boss/GPT-SoVITS}, ElevenLabs\footnote{https://elevenlabs.io/}, and CosyVoice~1.0~\cite{du2024cosyvoice}. By extending a public audio deepfake dataset with additional media degradations, the development condition provides an interpretable benchmark for controlled robustness analysis while remaining distinct from the final blind evaluation condition. Since both the original LlamaPartialSpoof dataset and the labeled development set are available after Phase~1, participants can perform detailed diagnostic analyses, including error inspection, score-distribution analysis, ablation studies, and hypothesis testing. As a result, the development set also serves as a meaningful benchmark for robustness analysis.

\begin{table*}[!t]
\centering
\caption{Bonafide speech sources used in the evaluation dataset.}
\label{tab:bonafide_sources}

\renewcommand{\arraystretch}{1.1}
\setlength{\tabcolsep}{6pt}

\begin{tabular}{lrrrrrrrr}
\hline
\textbf{Bonafide Source} & \textbf{en} & \textbf{en-sg} & \textbf{zh} & \textbf{zh-tw} & \textbf{ja} & \textbf{vi} & \textbf{Total} & \textbf{Subset \%} \\
\hline
Common Voice Scripted Speech \cite{commonvoice2020}    & 8,000 & 1,500 & 5,000 & 2,000 & 6,000 & 3,000 & 25,500 & 51.00\% \\
People's Speech \cite{galvez2021people}        & 7,000 & ---   & ---   & ---   & ---   & ---   &  7,000 & 14.00\% \\
IMDA~\cite{koh19_nsc_imda}                    & ---   & 2,000 & ---   & ---   & ---   & ---   &  2,000 &  4.00\% \\
MAGICDATA Mandarin Read Speech \cite{magicdata2019openslr68}      & ---   & ---   & 6,500 & ---   & ---   & ---   &  6,500 & 13.00\% \\
FormosaSpeech \cite{liao2020formosa} & ---   & ---   & ---   & 1,500 & ---   & ---   &  1,500 &  3.00\% \\
CPJD \cite{takamichi2018cpjd}           & ---   & ---   & ---   & ---   & 3,500 & ---   &  3,500 &  7.00\% \\
FOSD \cite{tran2020fosd}         & ---   & ---   & ---   & ---   & ---   & 4,000 &  4,000 &  8.00\% \\
\hline
Total                   & 15,000 & 3,500 & 11,500 & 3,500 & 9,500 & 7,000 & 50,000 & 100.00\% \\
\hline
\end{tabular}
\end{table*}

\begin{table*}[!t]
\centering

\begin{threeparttable}
\centering
\caption{Spoof speech sources used in the evaluation dataset.}
\label{tab:spoof_sources}

\renewcommand{\arraystretch}{1.1}
\setlength{\tabcolsep}{6pt}
\begin{tabular}{p{3.5cm}lrrrrrrrr}
\hline
\textbf{Text-to-Speech System} & \textbf{Type} & \textbf{en} & \textbf{en-sg} & \textbf{zh} & \textbf{zh-tw} & \textbf{ja} & \textbf{vi} & \textbf{Total} & \textbf{Subset \%} \\
\hline
iFlytek\tnote{1}    & Commercial  & 1,885 & ---   &   851 & --- &   313 & 1,474 & 4,523 &  8.58\% \\
Houshan\tnote{2}    & Commercial  & 1,678 & ---   &   989 & --- & 1,502 & ---   & 4,169 &  7.91\% \\
ElevenLabs\tnote{3} & Commercial  & 1,794 &   539 & 1,221 & 340 &   833 &   618 & 5,345 & 10.14\% \\
Cartesia\tnote{4}  & Commercial  & 1,609 & ---   &   809 & --- &   933 &   868 & 4,219 &  8.00\% \\
OpenAI\tnote{5}    & Commercial  &   876 & ---   & 1,415 & --- &   717 & 1,187 & 4,195 &  7.96\% \\ \hline
Chatterbox\tnote{6} & Open-source & 1,884 & 2,915 & 1,371 & --- &   939 & ---   & 7,109 & 13.48\% \\
CosyVoice~3.0~\cite{du2025_cosyvoice}  & Open-source & 1,770 & 2,688 & 1,098 & 348 & 1,371 & ---   & 7,275 & 13.80\% \\
Qwen3-TTS~\cite{hu2026qwen3}  & Open-source & 2,204 & ---   & 1,280 & 336 & 1,060 & ---   & 4,880 &  9.26\% \\
Fish Audio S2 Pro~\cite{liao2026fish} & Open-source & 1,232 &   569 & 1,592 & 654 & 1,074 &   841 & 5,962 & 11.31\% \\
Piper\tnote{7}      & Open-source & 2,991 & ---   &   608 & --- & ---   & 1,450 & 5,049 &  9.58\% \\
\hline
Total      & ---         & 17,923 & 6,711 & 11,234 & 1,678 & 8,742 & 6,438 & 52,726 & 100.00\% \\
\hline
\end{tabular}

\begin{tablenotes}[para,flushleft]
\footnotesize
% \item[1] https://global.xfyun.cn/
\item[1] https://www.xfyun.cn/services/online\_tts
% \item[2] https://www.volcengine.com/
\item[2] https://www.volcengine.com/product/tts
\item[3] https://elevenlabs.io/
\item[4] https://cartesia.ai
\item[5] https://openai.com/
\item[6] https://github.com/resemble-ai/chatterbox
\item[7] https://github.com/OHF-Voice/piper1-gpl
\end{tablenotes}
\end{threeparttable}
\end{table*}

\subsection{Evaluation Set}
The evaluation set was specifically developed as a new benchmark for robust audio deepfake detection under realistic multilingual and media-transformed conditions. Rather than maximizing language coverage, the benchmark focuses on several major language conditions relevant to the Asia-Pacific region: English (en), Singapore English (en-sg), Mandarin Chinese (zh), Taiwanese Mandarin (zh-tw), Japanese (ja), and Vietnamese (vi). Table~\ref{tab:language_distribution} summarizes the language and class distribution.
The final evaluation set contains 102,726 utterances, including 50,000 bona fide and 52,726 spoofed samples, resulting in a near-balanced overall class distribution. English is the largest language condition (32.05\%), followed by Mandarin (22.13\%), Japanese (17.76\%), and Vietnamese (13.08\%), while Singapore English and Taiwanese Mandarin account for 9.94\% and 5.04\% of the dataset, respectively. Although the overall dataset is near-balanced, class balance varies across language conditions, with Singapore English being spoof-heavy and Taiwanese Mandarin being bona fide-heavy. Since EER is threshold-based, language-dependent score distributions may affect global calibration.

The benchmark was deliberately designed to include multiple languages, bona fide sources, spoofing systems, and media-transformation conditions without enforcing uniform distributions across all factors. This better reflects practical deployment conditions, where detectors encounter uneven language distributions, heterogeneous recording sources, diverse synthesis systems, and unpredictable processing pipelines.

\subsubsection{Bona fide subset}
The bona fide subset contains 50,000 utterances collected from multiple speech corpora. Table~\ref{tab:bonafide_sources} summarizes the bona fide data sources. Most sources consist of read or natural speech recorded in relatively professional environments with a single dominant speaker. The largest source is Common Voice Scripted Speech, which contributes 25,500 utterances (51.00\% of the bona fide subset) spanning all target language conditions. Since Common Voice is a widely used public-domain speech dataset, many participants may have already incorporated portions of it into their training pipelines. Additional corpora were therefore included to provide broader linguistic and recording diversity across the benchmark..

\subsubsection{Spoof Subset}
The spoof subset contains 52,726 utterances generated using ten text-to-speech systems. Table~\ref{tab:spoof_sources} summarizes the source-level distribution. Most spoof sources contribute approximately 10\% of the spoof subset, resulting in a relatively balanced generator distribution. The spoof data were intentionally designed to be diverse, covering commercial, open-source, and research-oriented synthesis systems. For commercial services, speech was generated using voices provided directly by the respective platforms, whereas open-source systems employed different voice-cloning methodologies depending on the capabilities and design of each tool. The resulting diversity spans multiple synthesis families, architectures, languages, and generation styles, reducing the likelihood that detectors can rely on artifacts from a single generator and instead encouraging more robust and generalizable approaches. Not all synthesis systems supported all six language conditions, further increasing variability across spoof sources and language combinations.

\section{Media Transformation}
RADAR Challenge 2026 focuses on robust audio deepfake detection under realistic media-transformed conditions. Both the development and evaluation sets apply diverse transformation pipelines to bona fide and spoofed speech to simulate practical recording, transmission, and distribution workflows. Tables~\ref{tab:media_transform_dev} and~\ref{tab:media_transform_eval} summarize the development and evaluation pipelines, respectively. During data generation, each utterance passes through the full pipeline in sequence, where individual transformations are applied probabilistically according to their activation chances. As a result, a sample may undergo none, one, or multiple sequential transformations. In several stages, only one transformation is randomly selected from a related group, such as Opus, MP3, or AAC codec compression.

The development pipeline applies a moderate set of transformations intended to support controlled robustness analysis and system development. These include silence trimming, zero padding, room impulse response (RIR) reverberation, additive noise, background music, dynamic range compression, fade-in/fade-out effects, peak-level normalization, and codec compression using Opus and MP3. The transformation probabilities and parameter ranges were intentionally constrained to maintain relatively interpretable conditions while still introducing meaningful distribution shifts compared with clean speech.

In contrast, the evaluation pipeline introduces substantially greater diversity and variability in both transformation types and parameter ranges. In addition to the transformations used in the development set, the evaluation condition includes environmental sound events, bandwidth limitation, loudness normalization, resampling to 8~kHz, AAC compression, streaming dropout simulation, and speech perturbation. Several transformations also use more aggressive parameter ranges, such as stronger silence trimming, lower signal-to-noise ratios, and lower codec bitrates. Multiple transformations may be applied sequentially within a single processing chain, thereby simulating realistic media-processing pipelines encountered in practical deployment scenarios.

\begin{table}[t]
\centering
\caption{Media transformation pipeline used for the development set.}
\label{tab:media_transform_dev}
\footnotesize
\begin{tabular}{r|p{6.8cm}}
\hline
\textbf{Chance} & \textbf{Transformation} \\
\hline
50\% & Silence trimming (leading: 30--80\%, trailing: 30--90\%) \\
\hline
25\% & RIR reverberation (MIT RIR dataset \cite{traer2016statistics}) \\
\hline
50\% & Zero padding (leading: 0--250\,ms, trailing: 0--400\,ms) \\
\hline
\multirow{2}{*}{25\%} & Background noise (SNR: 15--25\,dB, MUSAN noise \cite{snyder2015musan}) \\
\cline{2-2}
 & Background music (SNR: 12--22\,dB, FMA small \cite{defferrard2017fma}) \\
\hline
15\% & Dynamic range compression \\
\hline
25\% & Fade-in/fade-out (fade-in: 20--80\,ms; fade-out: 50--150\,ms) \\
\hline
60\% & Peak-level normalization (target peak: $-12$ to $1$\,dBFS) \\
\hline
\multirow{2}{*}{50\%} & Opus codec (32--128\,kbps) \\
\cline{2-2}
 & MP3 codec (64--160\,kbps) \\
\hline
\end{tabular}
\end{table}

\begin{table}[t]
\centering
\caption{Media transformation pipeline used for the evaluation set.}
\label{tab:media_transform_eval}
\footnotesize
\begin{tabular}{r|p{6.8cm}}
\hline
\textbf{Chance} & \textbf{Transformation} \\
\hline
75\% & Silence trimming (leading \& trailing: 50--100\%) \\
\hline
20\% & RIR reverberation
(Aachen RIRs \cite{jeub2009binaural}, Simulated RIRs \cite{allen1979image}, Synthetic RIRs \cite{luong2024room}) \\
\hline
20\% & Zero padding (leading: 0--400\,ms, trailing: 0--500\,ms) \\
\hline
20\% & Background music (SNR: 8--20\,dB, FMA small \cite{defferrard2017fma}) \\
\hline
20\% & Sound Effect (SNR: 4--12\,dB, BSD10k sound effect \cite{anastasopoulou2025hierarchical}) \\
\hline
10\% & Bandwidth limitation (passband: 300--3400\,Hz;) \\
\hline
20\% & Dynamic range compression \\
\hline
20\% & Fade-in/fade-out (fade-in: 20--80\,ms; fade-out: 50--150\,ms) \\
\hline
100\% & Peak-level normalization (target peak: $-12$ to $1$\,dBFS) \\
\hline
20\% & Loudness normalization (target: $-18$ to $-14$\,LUFS) \\
\hline
10\% & Resampling to 8\,kHz (final output formatted at 16\,kHz) \\
\hline
\multirow{3}{*}{50\%} & Opus codec (24--96\,kbps) \\
\cline{2-2}
 & MP3 codec (48--128\,kbps) \\
\cline{2-2}
 & AAC codec (64--96\,kbps) \\
\hline
10\% & Streaming dropout simulation \newline
(frame: 10--40\,ms; dropout enter probability: 2\%--8\%) \\
\hline
10\% & Speech perturbation (speed factor: 0.95--1.05) \\
\hline
\end{tabular}
\end{table}

Real-world speech recordings often undergo multiple stages of processing before reaching an audio deepfake detector, including editing, normalization, codec compression, resampling, environmental interference, and room acoustics. These transformations can suppress spoofing artifacts or introduce confounding distortions; for example, codec compression may remove high-frequency synthesis artifacts, while reverberation and background interference can obscure temporal fine structure. Consequently, the challenge encourages the development of systems that rely on robust and generalizable evidence rather than narrow generator-specific artifacts.

\section{Baseline}

We provide a single baseline system based on the open-source SSL-AASIST anti-spoofing model~\cite{tak2022automatic}, which combines a wav2vec~2.0 frontend with the AASIST backend. Despite its moderate size (318M parameters), the model remains a strong and accessible system. We directly use the pretrained model released by the original authors and provide example evaluation scripts and score analysis~\cite{luong2025robust} for reproducibility and easy onboarding. Originally trained on ASVspoof 2019 LA~\cite{asvspoof2019database} with RawBoost-based augmentation, the baseline is not optimized for the multilingual and media-transformed conditions of RADAR 2026. It achieved 37.71\% EER on the development set and 42.6\% EER on the multilingual evaluation set. The baseline is intended as a reference system rather than a performance ceiling, and participants are encouraged to improve robustness through stronger augmentation, multilingual training data, calibration, model fusion, or alternative architectures. During Phase~1, teams were required to outperform the baseline on the development set in their final submission to qualify for the evaluation phase.

\section{Results}

\begin{figure}[t]
         \centering

         \includegraphics[width=\linewidth]{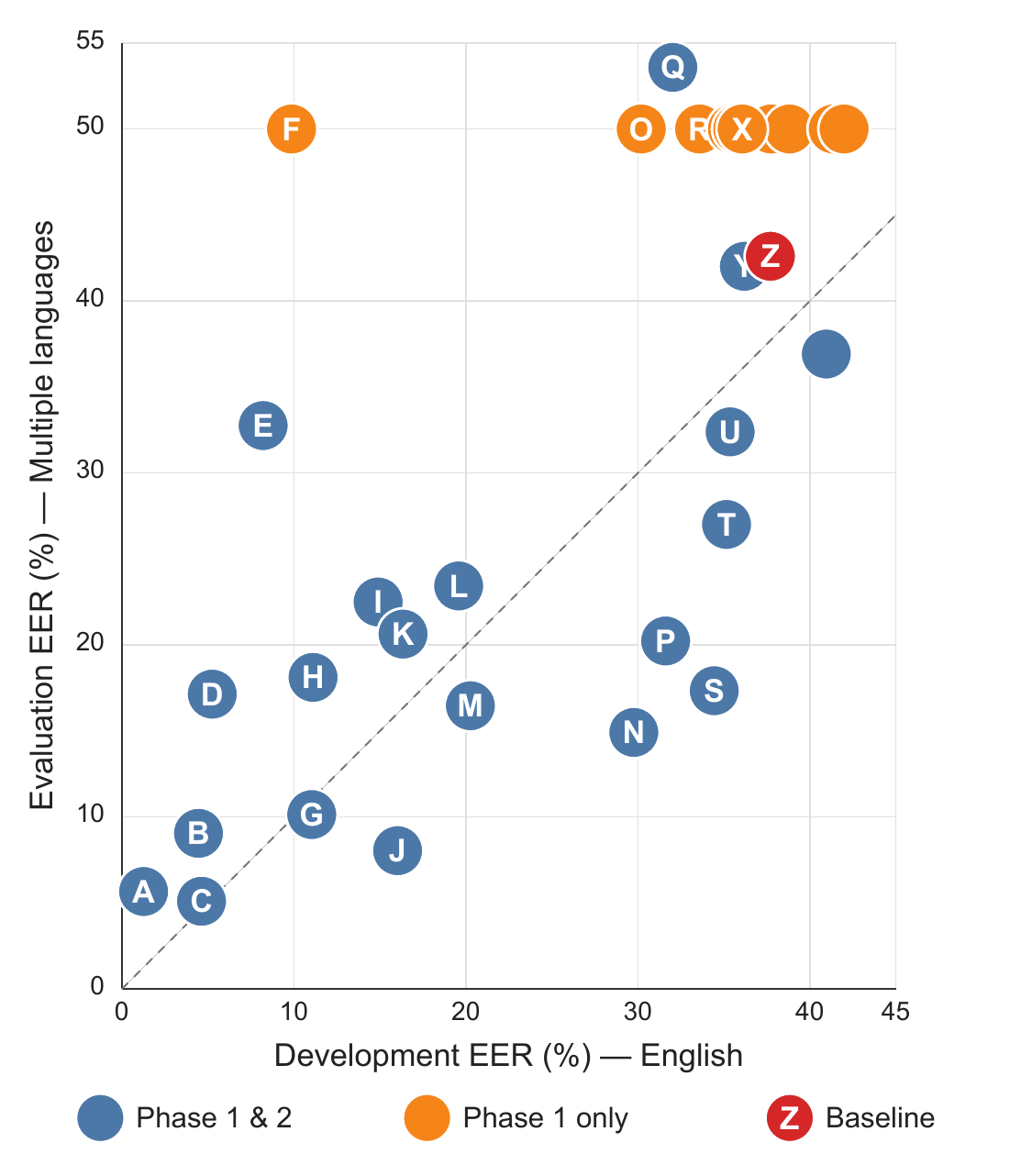}
         \vspace{-6mm}
         \caption{The EER results in Phase 1 and Phase 2 of the top 26 teams.}
    \vspace{-5mm}
    \label{fig:results}
\end{figure}
Figure~\ref{fig:results} shows the EER results of the top 26 teams. Each team is assigned an ID from A to Z according to its Phase~1 ranking, with Z corresponding to the baseline. Among them, 19 participated in the evaluation phase, while 7 did not for various reasons; these teams are assigned an evaluation EER of 50\% in the figure.
Team~A achieved the best Phase~1 result with 1.27\% and ranked second on the multilingual Phase~2 evaluation set with 5.67\%, whereas Team~C ranked third on the development set with 4.63\% and achieved the best Phase~2 result with 5.10\%. Team~B showed relatively stable performance, ranking second in Phase~1 with 4.46\% and fourth in Phase~2 with 9.05\%. In contrast, Team~J improved substantially from 10th place on the development set with 16.03\% to third place on the evaluation set with 8.04\%.

The leaderboards show that top-ranked systems achieved substantially lower EERs in both phases, indicating that robust performance under multilingual and media-transformed conditions is achievable with sufficiently strong approaches. Performance differences likely reflect variations in model design, training data, augmentation, and score calibration. The ranking changes between Phase~1 and Phase~2 also suggest that strong development performance does not always translate to robust blind evaluation performance under unseen conditions. Additional analyses and system-level comparisons will be reported when more information becomes available.

\section{Discussion}

\subsection{Limitations of Simulated Media Transformation}

Although the challenge introduces diverse media transformations to approximate realistic deployment conditions, the pipeline remains simulation-based and may not fully reflect real-world media processing~\cite{li2025measuring}. In practice, audio shared through communication platforms, social media, recording devices, and editing applications may undergo more complex or proprietary processing chains that are difficult to reproduce in a controlled benchmark~\cite{muller2025replaydf}. Another limitation is the assumption that transformed bona fide speech should always retain its original label. In practice, severe media processing may introduce artifacts that overlap with spoof-related characteristics, particularly under heavily degraded conditions. Nevertheless, for simplicity and consistency during the challenge, all transformed utterances retain their original labels.

\subsection{Limitations of the Current Analysis}

The current analysis is primarily based on aggregate leaderboard results and dataset statistics, and therefore does not yet provide detailed condition-level or system-level analysis. Since many participant system descriptions and training details are not yet publicly available, it remains difficult to determine which language conditions, media transformations, spoofing systems, model architectures, augmentation strategies, or training data contributed most to robustness and generalization performance. We therefore plan to conduct a more comprehensive follow-up analysis after additional participant reports and system-description papers become available.
In addition, the evaluation phase partly relies on a blind-evaluation assumption, since participants do not know the exact bona fide data sources used during evaluation; this blind condition will no longer be preserved once the evaluation set is publicly released.

\section{Conclusion}
This paper summarized the APSIPA RADAR Challenge 2026, a new benchmark for robust audio deepfake detection under realistic media transformations. The challenge includes an English development set for controlled analysis and a large-scale blind evaluation set spanning six language conditions and diverse media-processing pipelines. The results show that robust audio deepfake detection under multilingual and media-transformed conditions remains challenging, although several teams achieved promising performance despite the limited development timeline and blind evaluation setting. The benchmark is intended to support future research on media-robust audio deepfake detection, including transformation-aware training, multilingual modeling, calibration, and robustness evaluation under realistic deployment conditions. Both the development and evaluation sets will be publicly released after the conference, and future work will include more detailed condition-level and system-level analyses.

\section*{Acknowledgment}
Various AI-assisted tools and agents, including but not limited to ChatGPT, Claude, DeepSeek, Cursor, and Hermes Agent, were used during different stages of the challenge, including idea brainstorming, proposal writing, website development, dataset creation, email drafting, and paper preparation. The corresponding author takes full responsibility for all challenge-related content, materials, and conclusions.

% \begin{thebibliography}{1}

% \bibitem{1}
% G.~Eason, B.~Noble, and I.~N.~Sneddon, ``On certain integrals of
% Lipschitz-Hankel type involving products of Bessel functions,''
% \emph{Phil. Trans. Roy. Soc. London,} vol. A247, pp. 529-551, April
% 1955.

% \bibitem{2}
% J.~Clerk~Maxwell, \emph{A Treatise on Electricity and Magnetism,}
% 3$^{\rm rd}$ ed., vol. 2. Oxford: Clarendon, 1892, pp.68-73.

% \bibitem{3}
% I.~S.~Jacobs and C.~P.~Bean, ``Fine particles, thin films and exchange
% anisotropy,'' in \emph{Magnetism,} vol. III, G.T. Rado and H. Suhl,
% Eds. New York: Academic, 1963, pp. 271-350.

% \bibitem{4}
% K.~Elissa, ``Title of paper if known,'' unpublished.

% \bibitem{5}
% R.~Nicole, ``Title of paper with only first word capitalized,''
% \emph{J. Name Stand. Abbrev.,} in press.

% \bibitem{6}
% Y.~Yorozu, M.~Hirano, K.~Oka, and Y.~Tagawa, ``Electron spectroscopy
% studies on magneto-optical media and plastic substrate interface,''
% \emph{APSIPA Transl. J. Magn. Japan,} vol. 2, pp. 740-741, August 1987
% [\emph{Digests 9$^{\rm th}$ Annual Conf. Magnetics Japan,} p. 301,
% 1982].

% \bibitem{7}
% M.~Young, \emph{The Technical Writer's Handbook.} Mill Valley, CA:
% University Science, 1989.

% \end{thebibliography}

\printbibliography

@INPROCEEDINGS{llamapartialspof,
  author={Luong, Hieu-Thi and Li, Haoyang and Zhang, Lin and Lee, Kong Aik and Chng, Eng Siong},
  booktitle={Proc. ICASSP}, 
  title={LlamaPartialSpoof: An LLM-Driven Fake Speech Dataset Simulating Disinformation Generation}, 
  year={2025},
  volume={},
  number={},
  doi={10.1109/ICASSP49660.2025.10888070}}

@ARTICLE{asvspoof2021_summary,
  author={Liu, Xuechen and Wang, Xin and Sahidullah, Md and Patino, Jose and Delgado, Héctor and Kinnunen, Tomi and Todisco, Massimiliano and Yamagishi, Junichi and Evans, Nicholas and Nautsch, Andreas and Lee, Kong Aik},
  journal={IEEE/ACM Transactions on Audio, Speech, and Language Processing}, 
  title={{ASVspoof 2021: Towards spoofed and deepfake speech detection in the wild}}, 
  year={2023},
  volume={31},
  number={},
  pages={2507-2522},
  doi={10.1109/TASLP.2023.3285283}}

@article{asvspoof5,
title = {{ASVspoof 5: Design, collection and validation of resources for spoofing, deepfake, and adversarial attack detection using crowdsourced speech}},
journal = {Computer Speech \& Language},
volume = {95},
pages = {101825},
year = {2026},
issn = {0885-2308},
doi = {https://doi.org/10.1016/j.csl.2025.101825},
author = {Xin Wang and Héctor Delgado and Hemlata Tak and Jee-weon Jung and Hye-jin Shim and Massimiliano Todisco and Ivan Kukanov and Xuechen Liu and Md Sahidullah and Tomi Kinnunen and Nicholas Evans and Kong Aik Lee and Junichi Yamagishi and Myeonghun Jeong and Ge Zhu and Yongyi Zang and You Zhang and Soumi Maiti and Florian Lux and Nicolas Müller and Wangyou Zhang and Chengzhe Sun and Shuwei Hou and Siwei Lyu and Sébastien {Le Maguer} and Cheng Gong and Hanjie Guo and Liping Chen and Vishwanath Singh},
}

@article{asvspoof2019database,
author = {Wang, Xin and Yamagishi, Junichi and Todisco, Massimiliano and Delgado, Hector and Nautsch, Andreas and Evans, Nicholas and Sahidullah, Md and Vestman, Ville and Kinnunen, Tomi and Lee, Kong Aik and Juvela, Lauri and Alku, Paavo and Peng, Yu-Huai and Hwang, Hsin-Te and Tsao, Yu and Wang, Hsin-Min and Maguer, Sebastien Le and Becker, Markus and Henderson, Fergus and Clark, Rob and Zhang, Yu and Wang, Quan and Jia, Ye and Onuma, Kai and Mushika, Koji and Kaneda, Takashi and Jiang, Yuan and Liu, Li-Juan and Wu, Yi-Chiao and Huang, Wen-Chin and Toda, Tomoki and Tanaka, Kou and Kameoka, Hirokazu and Steiner, Ingmar and Matrouf, Driss and Bonastre, Jean-Fran{\c{c}}ois and Govender, Avashna and Ronanki, Srikanth and Zhang, Jing-Xuan and Ling, Zhen-Hua},
//doi = {10.1016/j.csl.2020.101114},
issn = {08852308},
journal = {Computer Speech {\&} Language},
month = {nov},
pages = {101114},
title = {{ASVspoof 2019: A large-scale public database of synthesized, converted and replayed speech}},
volume = {64},
year = {2020}
}

@INPROCEEDINGS{add2022,
  author={Yi, Jiangyan and Fu, Ruibo and Tao, Jianhua and Nie, Shuai and Ma, Haoxin and Wang, Chenglong and Wang, Tao and Tian, Zhengkun and Bai, Ye and Fan, Cunhang and Liang, Shan and Wang, Shiming and Zhang, Shuai and Yan, Xinrui and Xu, Le and Wen, Zhengqi and Li, Haizhou},
  booktitle={Proc. ICASSP}, 
  title={{ADD} 2022: the first Audio Deep Synthesis Detection Challenge}, 
  year={2022},
  volume={},
  number={},
  pages={9216-9220},
  doi={10.1109/ICASSP43922.2022.9746939}}

@inproceedings{add2023,
	title = {{ADD} 2023: the {Second} {Audio} {Deepfake} {Detection} {Challenge}},
	booktitle = {Proc. {IJCAI} {DADA Workshop}},
	author = {Yi, Jiangyan and Tao, Jianhua and Fu, Ruibo and Yan, Xinrui and Wang, Chenglong and Wang, Tao and Zhang, Chu Yuan and Zhang, Xiaohui and Zhao, Yan and Ren, Yong and Xu, Le and Zhou, Junzuo and Gu, Hao and Wen, Zhengqi and Liang, Shan and Lian, Zheng and Nie, Shuai and Li, Haizhou},
	month = may,
	year = {2023}
}

@inproceedings{kirill2025safe,
  title={SAFE: Synthetic Audio Forensics Evaluation Challenge},
  author={Kirill, Trapeznikov and Cummer, Paul and Pherwani, Pranay and Aslam, Jai and Davinroy, Michael and Bautista, Peter and Cassani, Laura and Stamm, Matthew},
  booktitle={Proc. ACM IH\&MMSEC Workshop},
  pages={174--180},
  year={2025}
}

@inproceedings{tak2022automatic,
  author={Hemlata Tak and Massimiliano Todisco and Xin Wang and Jee-weon Jung and Junichi Yamagishi and Nicholas Evans},
  title={{Automatic Speaker Verification Spoofing and Deepfake Detection Using Wav2vec 2.0 and Data Augmentation}},
  year={2022},
  booktitle={Proc. Odyssey 2022},
  pages={112--119}
}

@inproceedings{koh19_nsc_imda,
  title     = {{Building the Singapore English National Speech Corpus}},
  author    = {Jia Xin Koh and Aqilah Mislan and Kevin Khoo and Brian Ang and Wilson Ang and Charmaine Ng and Ying-Ying Tan},
  booktitle = {{Interspeech 2019}},
  pages     = {321--325},
}

@article{du2025_cosyvoice,
  title={{CosyVoice 3: Towards In-the-wild Speech Generation via Scaling-up and Post-training}},
  author={Du, Zhihao and Gao, Changfeng and Wang, Yuxuan and Yu, Fan and Zhao, Tianyu and Wang, Hao and Lv, Xiang and Wang, Hui and Shi, Xian and An, Keyu and others},
  journal={arXiv preprint arXiv:2505.17589},
  year={2025}
}

@inproceedings{zen2019libritts,
  title={{LibriTTS: A Corpus Derived from LibriSpeech for Text-to-Speech}},
  author={Zen, Heiga and Dang, Viet and Clark, Rob and Zhang, Yu and Weiss, Ron J and Jia, Ye and Chen, Zhifeng and Wu, Yonghui},
  booktitle={Interspeech 2019},
  pages={2638--2642},
  year={2019}
}

@InProceedings{casanova22a_yourtts,
  title = 	 {{{Y}our{TTS}: Towards Zero-Shot Multi-Speaker {TTS} and Zero-Shot Voice Conversion for Everyone}},
  author =       {Casanova, Edresson and Weber, Julian and Shulby, Christopher D and Junior, Arnaldo Candido and G{\"o}lge, Eren and Ponti, Moacir A},
  booktitle = 	 {ICML},
  pages = 	 {2709--2720},
  year = 	 {2022},
  editor = 	 {Chaudhuri, Kamalika and Jegelka, Stefanie and Song, Le and Szepesvari, Csaba and Niu, Gang and Sabato, Sivan},
  volume = 	 {162},
  series = 	 {Proceedings of Machine Learning Research},
  month = 	 {17--23 Jul},
  publisher =    {PMLR},
}

@inproceedings{casanova24_xtts2,
  title     = {{XTTS: a Massively Multilingual Zero-Shot Text-to-Speech Model}},
  author    = {Edresson Casanova and Kelly Davis and Eren Gölge and Görkem Göknar and Iulian Gulea and Logan Hart and Aya Aljafari and Joshua Meyer and Reuben Morais and Samuel Olayemi and Julian Weber},
  year      = {2024},
  booktitle = {{Interspeech 2024}},
  pages     = {4978--4982},
  issn      = {2958-1796},
}

@inproceedings{lim22_lj_jets,
  author = {Jungil Lim and Jinsu Ye and Seonman Chun and Sunhee Kim and Jaeyoung Cho},
  title = {{JETS: Jointly Training FastSpeech2 and HiFi-GAN for End to End Text to Speech}},
  year = 2022,
  booktitle = {Interspeech 2022},
  pages = {2338--2342},
}

@article{traer2016statistics,
  title={Statistics of natural reverberation enable perceptual separation of sound and space},
  author={Traer, James and McDermott, Josh H},
  journal={PNAS},
  volume={113},
  number={48},
  pages={E7856--E7865},
  year={2016},
  publisher={National Academy Sciences}
}

@inproceedings{jeub2009binaural,
  title={A binaural room impulse response database for the evaluation of dereverberation algorithms},
  author={Jeub, Marco and Schafer, Magnus and Vary, Peter},
  booktitle={2009 16th international conference on digital signal processing},
  pages={1--5},
  year={2009},
  organization={IEEE}
}

@article{allen1979image,
  title={Image method for efficiently simulating small-room acoustics},
  author={Allen, Jont B and Berkley, David A},
  journal={The Journal of the Acoustical Society of America},
  volume={65},
  number={4},
  pages={943--950},
  year={1979},
  publisher={Acoustical Society of America}
}

@inproceedings{luong2024room,
  title={Room impulse responses help attackers to evade deep fake detection},
  author={Luong, Hieu-Thi and Truong, Duc-Tuan and Lee, Kong Aik and Chng, Eng Siong},
  booktitle={Proc. SLT 2024},
  pages={623--629},
  year={2024},
  organization={IEEE}
}

@article{snyder2015musan,
  title={Musan: A music, speech, and noise corpus},
  author={Snyder, David and Chen, Guoguo and Povey, Daniel},
  journal={arXiv preprint arXiv:1510.08484},
  year={2015}
}

@inproceedings{defferrard2017fma,
  title={FMA: A Dataset For Music Analysis},
  author={Defferrard, Micha{\"e}l and Benzi, Kirell and Vandergheynst, Pierre and Bresson, Xavier},
  booktitle={18th International Society for Music Information Retrieval Conference},
  year={2017}
}

@inproceedings{anastasopoulou2025hierarchical,
  title={Hierarchical and multimodal learning for heterogeneous sound classification},
  author={Anastasopoulou, Panagiota and Dal R{\'\i}, Francesco Ardan and Serra, Xavier and Font, Frederic},
  booktitle={Proc. DCASE 2025},
  year={2025}
}

@inproceedings{commonvoice2020,
  author = {Ardila, R. and Branson, M. and Davis, K. and Henretty, M. and Kohler, M. and Meyer, J. and Morais, R. and Saunders, L. and Tyers, F. M. and Weber, G.},
  title = {Common Voice: A Massively-Multilingual Speech Corpus},
  booktitle = {Proceedings of the 12th Conference on Language Resources and Evaluation (LREC 2020)},
  pages = {4211--4215},
  year = 2020
}

@article{galvez2021people,
  title={The people's speech: A large-scale diverse english speech recognition dataset for commercial usage},
  author={Galvez, Daniel and Diamos, Greg and Ciro, Juan and Cer{\'o}n, Juan Felipe and Achorn, Keith and Gopi, Anjali and Kanter, David and Lam, Maximilian and Mazumder, Mark and Reddi, Vijay Janapa},
  journal={arXiv preprint arXiv:2111.09344},
  year={2021}
}

@misc{magicdata2019openslr68,
  author       = {{Magic Data Technology Co., Ltd.}},
  title        = {OpenSLR68: MagicData Mandarin Chinese Read Speech Corpus},
  year         = {2019},
  howpublished = {\url{http://www.imagicdatatech.com/index.php/home/dataopensource/data_info/id/101}},
  note         = {Accessed: 2019-05}
}

@inproceedings{liao2020formosa,
  title={Formosa speech recognition challenge 2020 and taiwanese across taiwan corpus},
  author={Liao, Yuan-Fu and Chang, Chia-Yu and Tiun, Hak-Khiam and Su, Huang-Lan and Khoo, Hui-Lu and Tsay, Jane S and Tan, Le-Kun and Kang, Peter and Thiann, Tsun-guan and Iunn, Un-Gian and others},
  booktitle={Proc. O-COCOSDA 2020},
  pages={65--70},
  year={2020},
  organization={IEEE}
}

@inproceedings{takamichi2018cpjd,
  title={CPJD corpus: Crowdsourced parallel speech corpus of Japanese dialects},
  author={Takamichi, Shinnosuke and Saruwatari, Hiroshi},
  booktitle={Proc. LREC 2018},
  year={2018}
}

@dataset{tran2020fosd,
  author    = {Tran, Duc Chung},
  title     = {{FPT Open Speech Dataset (FOSD) - Vietnamese}},
  year      = {2020},
  publisher = {Mendeley Data},
  version   = {V4},
  doi       = {10.17632/k9sxg2twv4.4}
}

@article{hu2026qwen3,
  title={Qwen3-TTS Technical Report},
  author={Hu, Hangrui and Zhu, Xinfa and He, Ting and Guo, Dake and Zhang, Bin and Wang, Xiong and Guo, Zhifang and Jiang, Ziyue and Hao, Hongkun and Guo, Zishan and others},
  journal={arXiv preprint arXiv:2601.15621},
  year={2026}
}

@article{liao2026fish,
  title={Fish Audio S2 Technical Report},
  author={Liao, Shijia and Wang, Yuxuan and Liu, Songting and Cheng, Yifan and Zhang, Ruoyi and Li, Tianyu and Li, Shidong and Zheng, Yisheng and Liu, Xingwei and Wang, Qingzheng and others},
  journal={arXiv preprint arXiv:2603.08823},
  year={2026}
}

@inproceedings{luong2025robust,
  title={Robust Localization of Partially Fake Speech: Metrics and Out-of-Domain Evaluation},
  author={Luong, Hieu-Thi and Rimon, Inbal and Permuter, Haim and Lee, Kong Aik and Chng, Eng Siong},
  booktitle={Proc. APSIPA ASC 2025},
  pages={2205--2210},
  year={2025},
  organization={IEEE}
}

@article{muller2025replaydf,
  title = {Replay Attacks Against Audio Deepfake Detection},
  author = {Nicolas Müller and Piotr Kawa and Wei-Herng Choong and Adriana Stan and Aditya Tirumala Bukkapatnam and Karla Pizzi and Alexander Wagner and Philip Sperl},
  journal={Interspeech 2025},
  year = {2025},
}

@article{li2025measuring,
  title={Measuring the robustness of audio deepfake detectors},
  author={Li, Xiang and Chen, Pin-Yu and Wei, Wenqi},
  journal={arXiv preprint arXiv:2503.17577},
  year={2025}
}

@article{du2024cosyvoice,
  title={Cosyvoice: A scalable multilingual zero-shot text-to-speech synthesizer based on supervised semantic tokens},
  author={Du, Zhihao and Chen, Qian and Zhang, Shiliang and Hu, Kai and Lu, Heng and Yang, Yexin and Hu, Hangrui and Zheng, Siqi and Gu, Yue and Ma, Ziyang and others},
  journal={arXiv preprint arXiv:2407.05407},
  year={2024}
}

@inproceedings{muller2024mlaad,
  title={Mlaad: The multi-language audio anti-spoofing dataset},
  author={M{\"u}ller, Nicolas M and Kawa, Piotr and Choong, Wei Herng and Casanova, Edresson and G{\"o}lge, Eren and M{\"u}ller, Thorsten and Syga, Piotr and Sperl, Philip and B{\"o}ttinger, Konstantin},
  booktitle={Proc. IJCNN 2024},
  pages={1--7},
  year={2024},
  organization={IEEE}
}

@article{wu2025sea,
  title={SEA-Spoof: Bridging The Gap in Multilingual Audio Deepfake Detection for South-East Asian},
  author={Wu, Jinyang and Hou, Nana and Pan, Zihan and Zhang, Qiquan and Bhupendra, Sailor Hardik and Mondal, Soumik},
  journal={arXiv preprint arXiv:2509.19865},
  year={2025}
}

@article{gao2025perturbed,
  title={Perturbed public voices (P$^2$V): A dataset for robust audio deepfake detection},
  author={Gao, Chongyang and Postiglione, Marco and Gortner, Isabel and Kraus, Sarit and Subrahmanian, VS},
  journal={arXiv preprint arXiv:2508.10949},
  year={2025}
}

@article{muller2025deepen,
  title={DeePen: Penetration Testing for Audio Deepfake Detection},
  author={M{\"u}ller, Nicolas and Kawa, Piotr and Stan, Adriana and Doan, Thien-Phuc and Jung, Souhwan and Choong, Wei Herng and Sperl, Philip and B{\"o}ttinger, Konstantin},
  journal={arXiv preprint arXiv:2502.20427},
  year={2025}
}

@article{kshirsagar2026investigating,
  title={Investigating the Impact of Speech Enhancement on Audio Deepfake Detection in Noisy Environments},
  author={Kshirsagar, Shruti and Avila, Anderson R and others},
  journal={arXiv preprint arXiv:2603.14767},
  year={2026}
}

\end{document}